# Blue Phosphorene Oxide: Strain-tunable Quantum Phase Transitions and Novel 2D Emergent Fermions

Liyan Zhu[†#*], Shan-Shan Wang[‡#], Shan Guan[‡], Ying Liu[‡], Tingting Zhang[†], Guibin Chen[†], and Shengyuan A. Yang[‡*]

[†]School of Physics and Electronic & Electrical Engineering, and Jiangsu Key Laboratory of Modern Measurement Technology and Intelligent Systems, Huaiyin Normal University, Huai'an, Jiangsu 223300, P. R. China
[‡]Research Laboratory for Quantum Materials, Singapore University of Technology and Design, Singapore 487372, Singapore

**ABSTRACT:** Tunable quantum phase transitions and novel emergent fermions in solid state materials are fascinating subjects of research. Here, we propose a new stable two-dimensional (2D) material, the blue phosphorene oxide (BPO), which exhibits both. Based on first-principles calculations, we show that its equilibrium state is a narrow-bandgap semiconductor with three bands at low energy. Remarkably, a moderate strain can drive a semiconductor-to-semimetal quantum phase transition in BPO. At the critical transition point, the three bands cross at a single point at Fermi level, around which the quasiparticles are a novel type of 2D pseudospin-1 fermions. Going beyond the transition, the system becomes a symmetry-protected semimetal, for which the conduction and valence bands touch quadratically at a single Fermi point that is protected by symmetry, and the low-energy quasiparticles become another novel type of 2D double Weyl fermions. We construct effective models characterizing the phase transition and these novel emergent fermions, and we point out several exotic effects, including super Klein tunneling, supercollimation, and universal optical absorbance. Our result reveals BPO as an intriguing platform for the exploration of fundamental properties of quantum phase transitions and novel emergent fermions, and also suggests its great potential in nanoscale device applications.

The field of two-dimensional (2D) materials has been attracting tremendous interest since the discovery of graphene in 2004.[1] It is now understood that many of graphene's extraordinary properties can be attributed to its distinct electronic band structure: the conduction and valence bands cross at isolated points in the reciprocal space, around which the low-energy electrons can be described by 2D ultra-relativistic Dirac fermions with a pseudospin-1/2.[2] This understanding has inspired active research to explore materials with unusual band crossings that could feature novel types of emergent fermions. Recent progress in this direction has been mostly on the three-dimensional (3D) bulk materials. Notably, the 3D Dirac and Weyl semimetals, recently discovered in a number of materials, have respectively four- and two-band crossing points, around which the low-energy quasiparticles resemble the 3D Dirac and Weyl fermions in high energy physics.[3-8] A six-band crossing point has also been observed in the 3D $Hg_{1-x}Cd_xTe$ crystal, for which the low-energy electrons are termed as Kane fermions.[9] From a fundamental perspective, with a reduced dimensionality, 2D fermions could behave drastically different from their 3D counterparts. From a practical point of view, 2D materials permit much easier control of their properties compared with 3D bulk materials, and they are better suited for nanoscale device applications. Hence it is of great interest to explore novel 2D materials which could host new types of fermionic excitations.

The family of 2D materials has undergone rapid expansion. One recent contender in the family that has been receiving significant attention is phosphorene — a single layer of black phosphorus (hereafter referred to as black phosphorene) with an armchair-like puckered lattice structure.[10-12] Unlike graphene as a zero-gap semimetal, black phosphorene has a nearly direct bandgap of ~1.5 eV,[12, 13] and compared with $MoS_2$ it enjoys a much improved carrier mobility.[10-12] Phosphorous is known to be stable in a large family of structures. Particularly, another 2D phosphorus allotrope known as blue phosphorene was proposed, which shares the same high stability as black phosphorene and structurally has a buckled honeycomb lattice more similar to graphene.[14] Very recently, single-layer blue phosphorene was successfully realized experimentally by the molecular beam expitaxy technique.[15] It is noted that blue phosphorene's bandgap (> 2 eV) is indirect and appears to be too big for electronics applications.[14] In terms of band topology, both (black and blue) phosphorenes are not so interesting: they are just conventional semiconductors with Schrodinger-type fermions. Moreover, they are prone to oxidize under ambient conditions, hence strategies such as encapsulation with other inert 2D materials have been developed to prevent the oxidation.[16]

On the other hand, controlled covalent functionalization including oxidation could be a powerful tool to engineer the physical and chemical properties of 2D materials. This has been proved in the realization of a wide range of graphene derivatives with versatile properties and in the generation of 2D topologically nontrivial states.[17-21] The phosphorene oxides have also been studied previously based on the black phosphorene structure, indicating the tunability of the band gaps with most of them being wide-gap insulators.[22-25]

Motivated by the great interest in search of nontrivial band crossings featuring novel fermionic excitations and by the experimental breakthrough in the synthesis of single-layer blue phosphorene,[15] here based on first-principles calculations, we propose a new 2D material which we term as blue phosphorene oxide (BPO), since its structure can be viewed as a fully oxidized single-layer blue phosphorene. We show that BPO is energetically and dynamically stable. At equilibrium state, BPO exhibits a tiny direct bandgap at the $\Gamma$ point. The low-energy spectrum involves three bands, of which two hole bands are degenerate at the $\Gamma$ point as dictated by the crystalline symmetry. We show that because of the buckled lattice structure, BPO is very soft and can sustain large strains. Strikingly, the bandgap closes at a small applied strain, in which the band edge states at the Γ point switch order without hybridization, corresponding to a semiconductor-to-semimetal quantum phase transition. Particularly, at the critical point of the phase transition, all the three low-energy bands cross at a single point on the Fermi level, leading to the emergence of a new type of 2D pseudospin-1 fermions. Such peculiar particles do not have a direct analogue in high energy physics, and they possess exotic properties such as super Klein tunneling and super-collimation. Beyond the phase transition, BPO becomes a symmetry-protected semimetal (SSM) with conduction and valence bands touching quadratically at a single Fermi point, featuring a new kind of 2D double Weyl fermions, and for which we find a remarkable property of a universal optical absorbance at low-frequency. Besides optical response, we argue that the electronic transport properties will also change dramatically across the phase transition. A low-energy effective model capturing the essential features of the phase transition and the emergent novel fermions have been constructed. Our result not only predicts a new member of the 2D materials family, more importantly, it reveals a fundamentally new type of quantum phase transition with two new types of 2D emergent fermions. The structural stability, the excellent mechanical property, and the highly tunable electronic and optical properties will make 2D BPO a promising material for nanoscale device applications.

Our first-principles calculations are based on the density functional theory (DFT) as implemented in Vienna ab-initio simulation package (VASP)[26-28]. The interactions between electrons and ions are modeled by the projector augmented wave (PAW) pseudopotential.[29, 30] The generalized gradient approximation (GGA) as parameterized by Perdew, Burke, and Ernzerhof (PBE)[31] is adopted to describe the exchange and correlation interactions between electrons. A vacuum space of 20 Å is used to avoid the interaction between periodic images. The energy cutoff is set to 400 eV for the plane-wave basis. The *k*-mesh of $25 \times 25 \times 1$ is adopted to sample the Brillouin zone. The cell parameters and ionic positions are fully optimized until the residual force on each atom is less than 0.01 eV/Å. For each strain, the atomic positions are fully relaxed without any constraint. The spin-orbit coupling strength in the system is very small due to the light elements involved. We have checked that the result obtained with spin-orbit coupling included shows negligible difference. Our main results are further confirmed by calculations with the hybrid functional method (HSE06)[32, 33] (Supporting Information). In the following presentation, we will focus on the GGA results.

The lattice structure of the proposed BPO is shown in Fig. 1(a, b). Without oxygen groups, it is reduced to blue phosphorene with a buckled honeycomb structure, in contrast to the black phosphorene which has a puckered armchair-like structure. The lattice of BPO has the space group symmetry of $D_{3d}^3$ (P-3m1), with two P and two O atoms in a primitive unit cell (see Fig. 1(a)). The lattice structure is similar to the experimentally realized graphane[34] and several other predicted 2D materials.[18, 20, 35, 36] The fully-relaxed structure has an equilibrium lattice constant of 3.68 Å. The equilibrium distances of P-P and P-O bonds are 2.38 Å and 1.48 Å, respectively. Compared with blue phosphorene (with lattice constant 3.33 Å and P-P bond length of 2.27 Å),[14] one observes that the in-plane parameters are slightly increased due to the presence of oxygen groups. Buckling means that the two P sublattices are on different atomic planes displaced along the perpendicular *z*-direction. The buckling can be characterized by the angle between the P-P and P-O bonds, which is 116.9º at equilibrium state (90º corresponds to the planar structure without buckling).

The structural stability can be inferred from the cohesive energy. In the previous study, it was shown that the blue phosphorene is as stable as black phosphorene by comparing their cohesive energies.[14] Here for BPO, the calculated cohesive energy is -8.288 eV per PO formula unit, which is also close to the value of black phosphorene oxide (-8.305 eV), indicating that the structural is almost equally stable as the black phosphorene oxide. We further check the dynamical stability of the BPO structure by calculating its phonon spectrum. The result is plotted in Fig. 1(d). One observes that there is no imaginary frequency in the phonon spectrum throughout the Brillouin zone, demonstrating that the BPO lattice is also dynamically stable (result for the 6% strained BPO is in the Supporting Information). In the vicinity of the Γ point, frequencies of out-of-plane acoustic phonons, known as the ZA phonons, show a parabolic dependence on the wave vector which is a characteristic feature of layered materials;[37-39] while linear dispersions are observed for the other two branches of in-plane acoustic phonons. Moreover, two branches of optical phonons, corresponding to rigid layer shear motion, emerge in the very low frequency region (~1.5 THz). An avoided crossing of these low-frequency optical phonons with acoustic phonons results in depressed dispersions of acoustic phonons. As a consequence, the sound speed of longitudinal acoustic phonons (~5.5 km/s) for BPO is smaller than that of the pristine blue phosphorene (~8.1 km/s).[14] The smaller sound speed in BPO indicates that the in-plane elastic stiffness of BPO should also be relatively small, as we show in the following.

We have calculated the strain-stress curve (see Supporting Information), which shows that BPO is within linear elastic regime up to 5% strain, and its elastic strain limit is very high (tensile strain limit is even beyond 30%). The elastic property of 2D materials is usually characterized by the in-plane stiffness constant, defined as $C = (1/S_0)(\partial^2 E_S/\partial \varepsilon^2)$, where $S_0$ is the equilibrium area of the cell, the strain energy $E_S$ is the energy difference between the strained and unstrained systems, and $\varepsilon$ is the in-plane uniaxial stain. From DFT calculation, we find that the stiffness constant is 49.5 N/m for BPO. This value is much smaller compared with other typical 2D materials, including graphene ($\sim 340 \pm 40$ N/m),[40] MoS$_2$ ($\sim$140 N/m),[41] and BN ($\sim$267 N/m),[42] indicating that the material is softer. Physically, this is due to the buckled structure which allows lattice deformation in the *z*-direction without much cost of energy. We find that the buckling angle changes from 118.5º at -4% strain to 115.4º at +4% strain. The great flexibility of BPO would facilitate the application of external strain for tuning its properties.

The most interesting feature of BPO lies in its electronic property. The electronic band structure for BPO at its equilibrium state is shown in Fig. 2(a), demonstrating a semiconducting phase with a small direct bandgap ~0.09 eV at the Γ point. One notes that the band edges involve three low-energy bands: a non-degenerate conduction band, and two valence (heavy-hole and light-hole) bands that are degenerate at the Γ point. The degeneracy of the two valence bands is dictated by symmetry: the two states at the valence band top correspond to the two-dimensional irreducible representation $E_g$ of the $D_{3d}$ point group at the Γ point (whereas the state at the conduction band bottom is of the one-dimensional $A_{2u}$ representation). These symmetry characters play a crucial role in the properties of BPO, as will be discussed in a while. By projecting the band states to atomic orbitals, one observes that the conduction band edge is mainly of the *s* and $p_z$ orbital character at the P sites, while the states at the valence band top mainly consist of Oxygen $p_x$ and $p_y$ orbitals. The charge density distributions for these states are plotted in Fig. 2(b, c).

Due to the small size of the bandgap, one expects that a small external perturbation may close the gap and drive a phase transition. In the following, we focus on the perturbation from the applied biaxial strain. The band structures for three representative strains of -4%, 0.6%, and +4% are shown in Figs. 3(a-c), respectively. Indeed, we find that the direct bandgap at the Γ point is closed at a small tensile strain of 0.6% (Fig. 3(b)), marking the critical point for a phase transition (the critical strain is 3.3% on HSE06 level, see Supporting Information). For tensile strains above the critical strain, BPO becomes a semimetal with the conduction and valence bands touching at a single Fermi point (Fig. 3(c)). On the opposite side, the system is a semiconductor with an increasing bandgap as the lattice is compressed (Fig. 3(a)).

Remarkably, one notes that because of the preserved crystalline symmetry, the states at the $\Gamma$ point maintain their symmetry characters, hence the $A_{2u}$ and $E_g$ states cannot hybridize: the two levels must directly cross each other during the phase transition process. Therefore, the strain-induced phase transition here also represents a quantum phase transition. This is in contrast to the usual scenario of level switching in which a direct crossing is avoided due to hybridization. In Fig. 3(d), we plot the energies of the nondegenerate $A_{2u}$ state and the doubly-degenerate $E_g$ states (referenced to the vacuum level) as a function of strain. Indeed, they cross each other at the critical strain of 0.6%, during which the band ordering is switched. Furthermore, beyond the critical point into the semimetal phase, the $E_g$ doublet is above the $A_{2u}$ state in energy. Due to the band filling, the Fermi level must directly cut through the $E_g$ states (see Fig. 3(c)). In this phase, the system remains a semimetal with a single Fermi point, and as mentioned, this band-crossing Fermi point is symmetry-protected: the degeneracy at this point cannot be lifted as long as the three-fold rotational symmetry is preserved. Thus the obtained semimetal phase is a novel kind of SSM phase, and the corresponding phase transition is a semiconductor-to-SSM quantum phase transition.

Accompanying the phase transition, there are interesting fermions emerging near the band-crossing points. Exactly at the critical point of the transition, the three low-energy bands cross at a single triply-degenerate point (at $\Gamma$) (see Figs. 3(b) and 4(a)). The low-energy quasiparticles are hence a novel type of three-component fermions (we will show in a while that they correspond to 2D pseudospin-1 fermions). After the transition to the SSM phase, two bands cross quadratically at a single doubly-degenerate point at the Fermi level (see Figs. 3(c) and 4(b)). The low-energy quasiparticles become two-component 2D double Weyl fermions.

The features of emergent fermions are characterized by the *k·p* effective model expanded around the band-crossing points. Using the three basis of the $A_{2u}$ and $E_g$ states, we construct the *k·p* effective model subjected to the $D_{3d}$ symmetry constraint. The three generators of the point group can be taken as the three-fold rotation, the vertical mirror plane, and the inversion. Plus the requirement of time reversal symmetry, the obtained *k·p* Hamiltonian expanded around the Γ point up to *k*-quadratic terms takes the general form of

$$H(\boldsymbol{k}) = (C + Dk^2)\mathbb{I}_{3\times 3} + \begin{bmatrix} -M + Bk^2 & -iAk_x & iAk_y \\ iAk_x & 0 & 0 \\ -iAk_y & 0 & 0 \end{bmatrix}, \qquad (1)$$

where *A, B, C, D,* and *M* are band parameters, *k* is the magnitude of the 2D wave vector $\boldsymbol{k} = (k_x, k_y)$, and $\mathbb{I}_{3\times 3}$ is the 3 × 3 identity matrix. The model parameters are obtained by fitting with the DFT band structure. The variation of the parameters versus strain is plotted in Fig. 5(a). One observes that the parameter *M* gives the energy difference between the $A_{2u}$ and $E_g$ states at the Γ point, hence the phase transition is marked by the sign change of *M*: in the semiconductor phase, *M*<0, the $A_{2u}$ state is above the $E_g$ doublet; whereas in the SSM phase, *M*>0, the ordering of the $A_{2u}$ and $E_g$ states are switched (see the inset of Fig. 5(a)).

Exactly at the critical point, we have *M*=0. Interestingly, if we only keep the *k*-linear terms, the model in Eq.(1) can be re-written in a more compact form with

$$H_{\text{spin-1}}(\bm{k}) = A\bm{k}\cdot\bm{S}, \qquad (2)$$

where

$$S_x = \begin{bmatrix} 0 & -i & 0 \\ i & 0 & 0 \\ 0 & 0 & 0 \end{bmatrix}, \quad S_y = \begin{bmatrix} 0 & 0 & i \\ 0 & 0 & 0 \\ -i & 0 & 0 \end{bmatrix}, \quad S_z = \begin{bmatrix} 0 & 0 & 0 \\ 0 & 0 & i \\ 0 & -i & 0 \end{bmatrix}$$

are the spin-1 matrices satisfying the algebra of angular momentum $[S_i, S_j] = i\epsilon_{ijk} S_k$. Since the emergent spin degree of freedom is not from the real electron spin (which is a dummy degree of freedom here), the low-energy quasiparticles are the 2D pseudospin-1 fermions. Such fermion is helical, in the sense that it has a well-defined helicity of ±1 and 0 corresponding to the eigenvalues of the helicity operator $\bm{k}\cdot\bm{S}/k$. Quasiparticles with pseudospin-1 have been studied mostly in the cold atom context and recently begin to appear in solid state systems.[43-46] It has been predicted that such particles can exhibit exotic transport effects including super Klein tunneling and supercollimation. Super Klein tunneling means that perfect transmission can occur for almost all the incident angles in the Klein tunneling process of pseudospin-1 particles.[43] And supercollimation refers to a guided unidirectional transport of a chiral (helical) particle in a periodic potential, regardless of its initial direction of motion.[47] Thus BPO provides a new 2D solid state platform to probe these fascinating effects associated with pseudospin-1 fermions.

After transition into the SSM phase, the band with $A_{2u}$ state is pushed below the valence band top, and the low-energy physics are manifested by the quasiparticles around the quadratic band-touching point from the two $E_g$ states. The *k·p* effective model can be obtained by projecting model (1) onto the $E_g$ subspace. After straightforward calculation and a unitary transformation, the effective model can be written as

$$\mathcal{H}(\bm{k}) = (C + D_1 k^2)\mathbb{I}_{2\times 2} + D_2 \begin{bmatrix} 0 & (k_x - ik_y)^2 \\ (k_x + ik_y)^2 & 0 \end{bmatrix}, \qquad (3)$$

where $D_2 = A^2/2(M-C)$, and $D_1 = D + D_2$. Without the diagonal term, the model resembles that of the AB-stacked bilayer graphene,[2, 48] which features a Berry phase of $2\pi$, doubling that of the 2D Weyl point as in single-layer graphene. Hence such band-touching point can be termed as a 2D double Weyl point.

Of the many possible interesting effects associated with 2D double Weyl fermions, we point out a universal optical absorbance of the SSM phase for the low-frequency light. Optical absorbance $A(\omega) = W_a/W_i$ is the ratio between the energy flux $W_a$ absorbed by the material and the energy flux $W_i$ of incident light with frequency $\omega$.[49] At low frequencies, when $\hbar\omega < |M|$, optical transitions are only between the two $E_g$ bands around the Fermi level. Then the absorbance can be evaluated using the effective model (3) by noticing that $W_a = \eta\hbar\omega$ with $\eta$ the absorption rate which can be simply calculated using the Fermi's golden rule: $\eta = 2 \times \frac{2\pi}{\hbar}\sum_{\boldsymbol{k}}|\mathcal{M}(\boldsymbol{k})|^2 \delta[E_c(\boldsymbol{k}) - E_v(\boldsymbol{k}) - \hbar\omega]$, where $\mathcal{M}$ is the optical coupling matrix element, $E_c$ and $E_v$ here refer to the band energies of the conduction and valence bands respectively, and the factor of 2 is from the real spin degeneracy. Straightforward calculation shows that $A(\omega) = \pi\alpha \cong 2.3\%$ with $\alpha = \frac{e^2}{\hbar c} \cong 1/137$ being the fine structure constant. This absorbance is universal in the sense that it only consists of the fundamental constants and is independent of the light frequency. Note that this universal value is valid in the frequency range with $\hbar\omega < |M|$; above this range, transitions from the $A_{2u}$ valence band becomes possible hence the absorbance will deviate from this $\pi\alpha$ value. The universal absorbance of 2.3% has been experimentally shown in single-layer graphene.[49] In comparison, here the band dispersion is quadratic instead of linear, giving rise to an extra factor of two in the absorption rate; however, the BPO bands only have a single valley whereas graphene has two valleys, thus the combined effect is such that BPO in the SSM phase shares the same universal absorbance as the single-layer graphene. We also checked this directly from the DFT calculations.[50] In calculating the optical property, we use a refined *k*-mesh of $81 \times 81 \times 1$. The low-energy part of the absorbance is calculated using a further refined non-uniform *k*-mesh with ~40,000 *k*-points in a region of $0.08 \times 2\pi/a$ around the $\Gamma$ point. The obtained absorbance as a function of frequency is plotted in Fig. 5(b), corresponding to the SSM state with a +4% strain. One clearly observes that the absorbance approaches the universal value of $\pi\alpha$ at frequencies below $|M|\sim 0.5$ eV.

This unusual optical absorption property can be used as a powerful probe of the quantum phase transition: on the semiconductor side, there is an absorption gap and the absorbance is generally frequency dependent; on the SSM side, the absorption does not have a gap and approaches a universal value of $\pi\alpha$ at low frequencies (see Fig. 5(b) for the comparison). The phase transition will also manifest in the carrier mobility for the *p*-doped case. In the semiconductor phase, both heavy-hole and light-hole bands contribute to the transport, whereas in the SSM phase, only the heavy-hole band contributes at small doping levels. As a result, one expects that the average hole mobility will have a large decrease across the semiconductor-to-SSM phase transition.

Before concluding, a few remarks are in order. First, while the 2D double Weyl point in the SSM phase of BPO is protected by the $C_3$ symmetry, the triply-degenerate point at the critical strain is not protected by symmetry or topology. Indeed, this is why we can have a tunable quantum phase transition. We mention that another possible way to tune the bandgap without breaking the rotational symmetry is by applying a vertical electric field. When the $C_3$ symmetry is broken, the degeneracy at the double Weyl point is no longer protected. In the SSM phase, this would either open a finite bandgap or split the double Weyl node into two linearly dispersive

Weyl nodes. We find that with an additional uniaxial strain in the SSM phase, the double Weyl node in BPO is always gapped (see Supporting Information).

Second, the main findings here, including the phase transition, the novel emergent fermions and their unusual properties are mostly determined by symmetry. Hence they are quite robust. For example, different functionals give qualitatively the same result (see HSE result in the Supporting Information). And in the presence of disorders, inferring from the previous studies in disordered graphene[51] and topological crystalline insulators,[52] as long as the symmetry is preserved on average, we expect that the effects predicted here will still survive. This also suggests that the underlying mechanism is general and possible to extend to other materials with similar symmetry. The result here will definitely motivate and provide useful guidance for searching these exciting phenomena in other 2D materials as well.

Third, we have also examined the 2D materials AsO, PS, and AsS with the same type of lattice structure by replacing the elements from the same groups in the periodic table. We find that for these materials, unlike BPO, at equilibrium state there are only two $E_g$ bands close to the Fermi level, while the $A_{2u}$ band is pushed away (see Supporting Information). For AsO, it is in the SSM phase with a single double Weyl point at the Fermi level. For PS and AsS, the conduction band at the M point of the Brillouin zone is lower in energy than that of the $\Gamma$ point, such that they become metals with a finite Fermi surface.

Finally, the technique of strain engineering on 2D materials has been well developed. For example, it can be achieved by attaching the 2D material to a flexible substrate and then applying mechanical strain to the substrate.[53, 54] If one uses a piezoelectric material as substrate, the strain may also be controlled by electric means.[55] On the material side, the blue phosphorene has recently been realized by epitaxial growth on Au(111) substrate.[15] The next step would be to isolate the blue phosphorene layer from the metallic substrate. Although challenging, we note that a technique has recently been developed for isolating silicene from the Ag(111) substrate.[56] After transferring the blue phosphorene to inert substrate with holes or trenches, one can then oxidize it as like the process for fabricating graphane.[34] In addition, for black phosphorene, it is experimentally demonstrated that the oxidation process can be well controlled, e.g., through the assistance of laser.[57] This may also be useful for the realization of BPO.

In conclusion, a new 2D material, the blue phosphorene oxide, has been proposed, which exhibits a range of fascinating properties. We demonstrate its energetic and dynamic stability. Its elastic property shows that the material is quite soft and can be easily tuned by strain. Its band structure involves three bands near the Fermi level. Most strikingly, we reveal a semiconductor-to-SSM quantum phase transition in BPO that can be readily tunable by strain. Across the phase transition, there are two novel types of fermions emerging at low energy: the 2D pseudospin-1 fermions at the critical point and the 2D double Weyl fermions in the SSM phase. We formulate effective models characterizing their features. The pseudospin-1 fermions could exhibit exotic super Klein tunneling and supercollimation effects, and we point out a universal optical absorbance associated with the double Weyl fermions. As a result, optical and transport measurements on BPO could provide distinct signatures for probing the phase transition. The proposed BPO material thus offers an intriguing platform not only for the fundamental exploration of novel emergent fermions beyond the Dirac and Weyl paradigm, but also for many possible electronic and optical applications at nanoscale.

AUTHOR INFORMATION

**Corresponding Author**

*Email: lyzhu@hytc.edu.cn

*Email: shengyuan_yang@sutd.edu.sg

**Author Contributions**

#L.Z. and S.-S.W. contributed equally to this work.

ACKNOWLEDGMENTS

We thank D.L. Deng for valuable discussions. This work was supported by National Natural Science Foundation of China (Grant No. 11504122 and 11547192). SSW, YL, SG, and SAY acknowledge support by funding Singapore MOE Academic Research Fund Tier 1 (SUTD-T1-2015004). LZ and TZ thank the funding support from Jiangsu Key Laboratory of Modern Measurement Technology and Intelligent Systems and Natural Science Foundation of the Higher Education Institutions of Jiangsu Province (Grant No. 15KJB140001).

# Figures

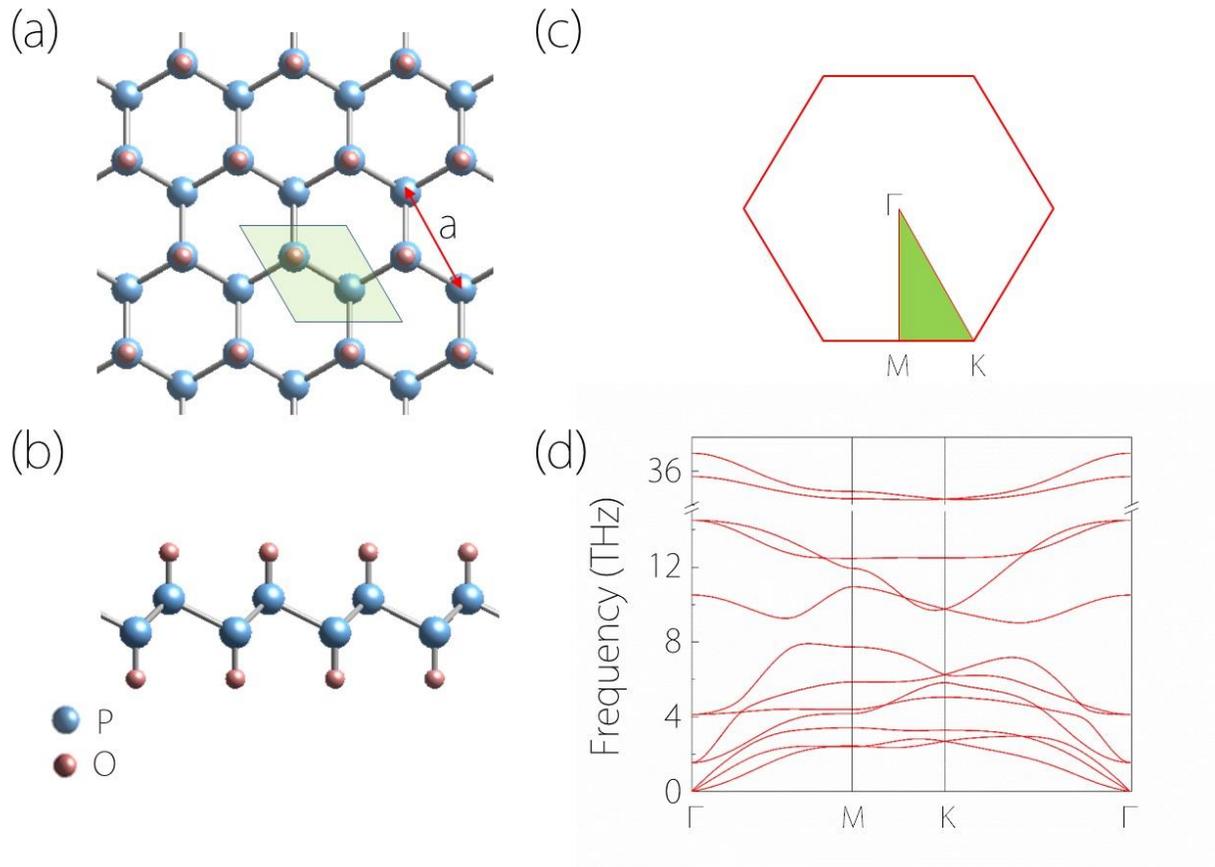

**Figure 1.** (a) Top and (b) side views of a 2D single-layer blue phosphorene oxide (BPO). In (a), the green shaded region indicates the primitive unit cell. $a$ is the lattice parameter. (c) First Brillouin zone with high-symmetry points labeled. (d) Phonon dispersion of BPO. A $2 \times 2$ supercell is used in the calculation.

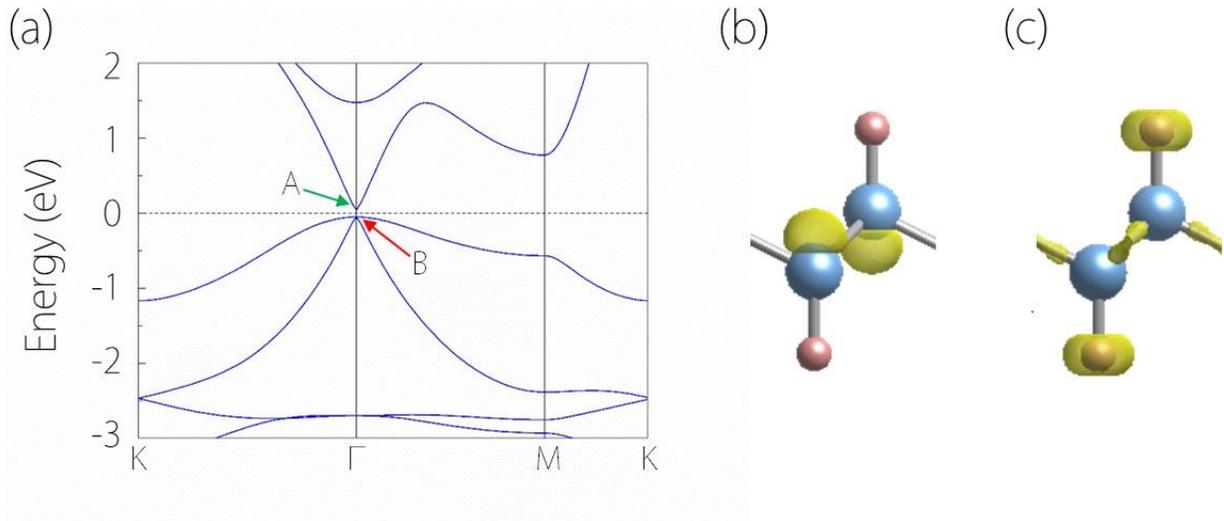

**Figure 2.** (a) Electronic band structure for BPO at equilibrium state. In the figure, A labels the non-degenerate state at the conduction band bottom, and B labels the doubly-degenerate state at the valence band top, both located at the $\Gamma$ point. (b) and (c) are the charge density distribution corresponding to states A and B, respectively.

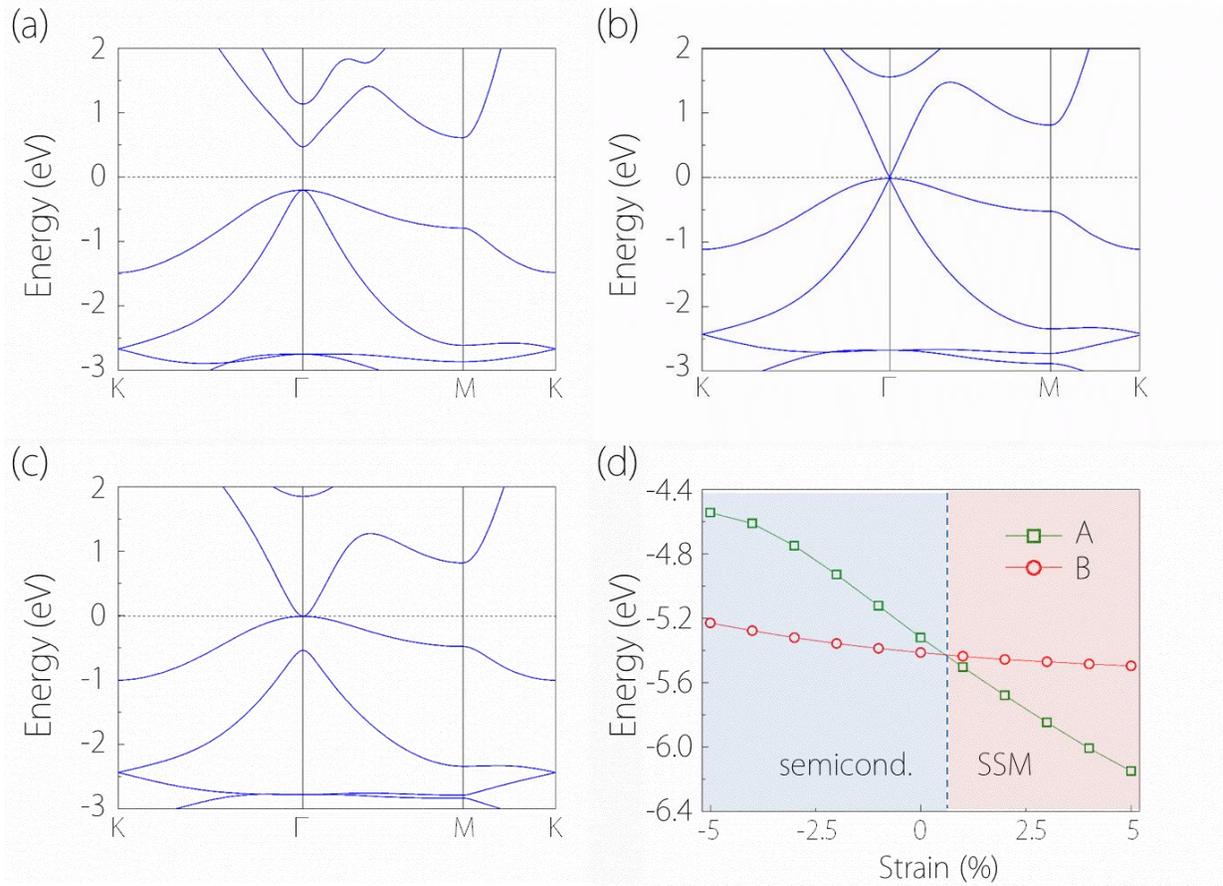

**Figure 3.** Electronic band structures of BPO under applied strain of (a) -4%, (b) +0.6%, and (c) +4%. (d) Energies of states A and B (indicated in Fig. 2(a)) versus strain. The crossing of the two corresponds to the quantum phase transition between a semiconductor phase and a symmetry-protected semimetal (SSM) phase. In (d), the energies are referenced to the vacuum level.

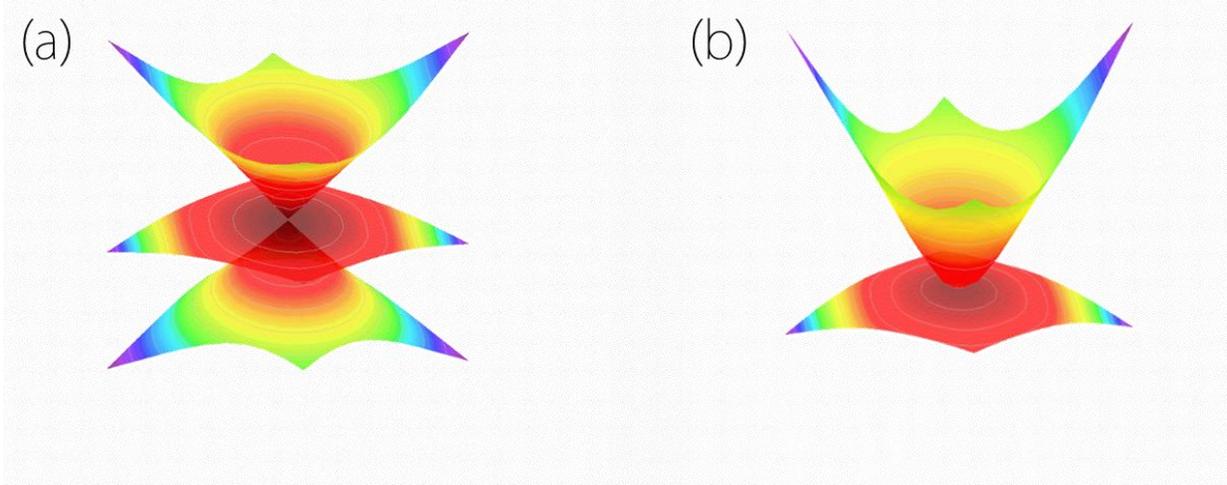

**Figure 4.** Energy dispersion around the band crossing point at Fermi level for (a) critical strain +0.6% and (b) +4% strain. The two figures correspond to the band structures in Fig. 3(b) and 3(c) respectively. The low-energy quasiparticles around the crossing point are described by (a) 2D pseudospin-1 fermions and (b) 2D double Weyl fermions.

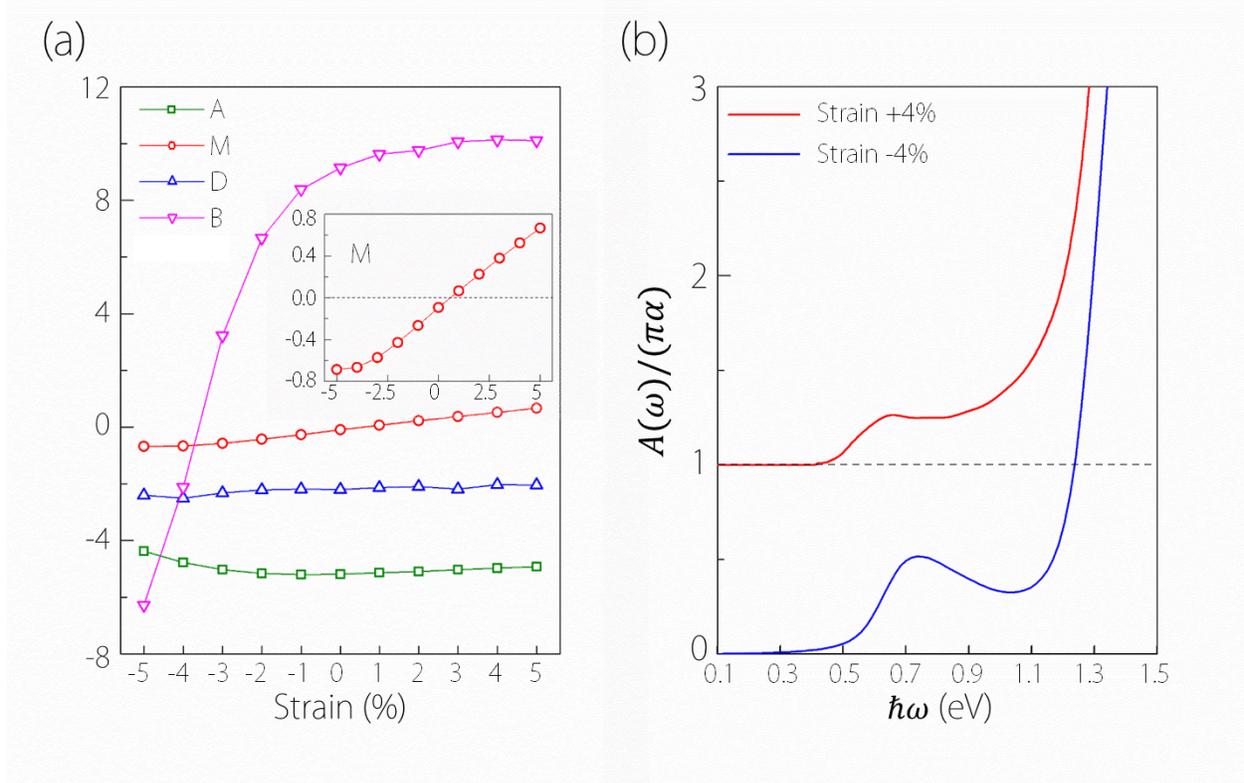

**Figure 5.** (a) Parameters in the low-energy effective model (Eq.(1)) as functions of strain. Here $M$ has the unit of eV, $B$ and $D$ have the unit of eV·Å$^2$, and $A$ has the unit of eV·Å. The inset is an enlarged figure for the variation of parameter $M$. The parameter $C$ in the model is an overall energy shift hence is not shown. (b) Optical absorbance as a function of frequency. The two curves are for the -4% strain in the semiconductor phase and for the +4% strain in the SSM phase, respectively. In the SSM phase, the system shows a universal optical absorbance of $\pi\alpha$ at low frequencies.